\documentclass[prl,twocolumn,nofootinbib]{revtex4}

\usepackage{epsfig}
\usepackage{amssymb}
\usepackage[tbtags]{amsmath}
\usepackage{upref}
\usepackage{float}
\usepackage{wrapfig}
\usepackage{subfigure}
\usepackage{longtable}
\newcommand{\crosssec}{\sigma_{p}^{\rm SI}}
\newcommand{\crosssecsd}{\sigma_{p}^{\rm SD}}

\newcommand{\ccrosssec}{\sigma_{p}^{\rm SI}}
\newcommand{\ccrosssecsd}{\sigma_{p}^{\rm SD}}

\def\bone{B^{(1)}}

\def\lsim{\raise0.3ex\hbox{$\;<$\kern-0.75em\raise-1.1ex\hbox{$\sim\;$}}}
\def\gsim{\raise0.3ex\hbox{$\;>$\kern-0.75em\raise-1.1ex\hbox{$\sim\;$}}}

\begin{document}

\title{WIMP identification through a combined measurement of axial and
  scalar couplings}
\author{G.~Bertone\,$^{1,2}$
      D.G.~Cerde\~no\,$^{3}$
      J.I.~Collar\,$^{4}$, and
      B.~Odom\,$^{4}$}
\affiliation{${^1}$ INFN, Sezione di Padova, Via Marzolo 8, Padova
  I-35131,
  Italy}  
\affiliation{${^2}$Institut d'Astrophysique de Paris, UMR 7095-CNRS
  Universit\'e Pierre et Marie Curie, 98bis boulevard Arago, 75014
  Paris, France}
\affiliation{${^3}$ Departamento de F\'{i}sica Te\'{o}rica C-XI \&
  Instituto de F\'{i}sica Te\'{o}rica C-XVI, Universidad Aut\'{o}noma
  de Madrid, Cantoblanco, E-28049 Madrid, Spain}
\affiliation{${^4}$ Enrico Fermi Institute and Kavli Institute for
  Cosmological Physics,\\ University of Chicago, IL 60637, USA}

\begin{abstract}
  We study the prospects for detecting Weakly Interacting Massive 
  Particles (WIMPs), in a number of phenomenological scenarios, with a 
  detector composed of a target simultaneously sensitive to both
  spin-dependent and spin-independent couplings, as is the case of
  COUPP (Chicagoland Observatory for Underground Particle Physics).
  First,
  we show that sensitivity to both couplings optimizes chances of
  initial  
  WIMP detection.  Second, we demonstrate that in
  case of detection, comparison of the signal on two complementary
  targets, such as in COUPP CF$_{3}$I and C$_{4}$F$_{10}$ bubble
  chambers, allows a significantly more precise determination of the
  dark matter axial and scalar couplings. This strategy would provide
  crucial 
  information on the nature of the WIMPs, and possibly allow 
  discrimination between neutralino and Kaluza-Klein dark matter.
\end{abstract}

\maketitle
\paragraph{Introduction.}
A variety of astrophysical and cosmological observations provide
convincing evidence that the matter budget of the Universe is
dominated by {\it Dark Matter} (DM),  made of some new,
yet undiscovered, particles that interact weakly or less-than-weakly
with those of the Standard Model (SM). The fact that some
well-motivated 
extensions of the SM, such as Supersymmetry (SUSY) and theories with
extra-dimensions,
{\it naturally} provide excellent DM candidates,
has attracted the interest of the particle physicists community, and
many current and upcoming searches are planned to tackle the question
of the nature and the properties of DM particles (for recent reviews,
see, e.g.,
Refs.~\cite{Jungman:1995df,Bergstrom:2000pn,mireview,Bertone:2004pz}).

DM can be
searched for {\it directly}, as DM particles passing through the
Earth interact inside large detectors. The field of direct searches is 
well-established, with many experiments currently operating 
or planned. Direct
detection relies on one of two modes of interaction with target
nuclei.  Scalar, or so-called {\it spin-independent}, coupling
describes coherent interactions of DM with the entire nuclear mass.
Axial, or so-called {\it spin-dependent}, coupling describes
interaction of DM with the spin-content of the nucleus. Overall,
more attention has been given to interpretation of direct detection
results in terms of the scalar interaction, and most experimental
efforts have focused on using heavy-nucleus targets which enhance
the scalar-interaction scattering rate. However, as detailed below,
it is generally not known whether first direct detection of DM
particles is more likely to occur via scalar or axial interactions.
Furthermore, determination of the nature of DM parameters will
require quantification of both types of interaction by measurement
of the scattering cross-section on multiple target nuclei.
Finally, spin-dependent couplings are also of
special interest since one of the most promising indirect searches
aim at the detection of high energy neutrinos from DM annihilations
at the center of the Sun, where DM would accumulate precisely due to
spin-dependent interactions with the nuclei of the Sun.

In this letter we concentrate the discussion around a new
experiment, COUPP (Chicagoland Observatory for Underground Particle
Physics), which exploits an old technique, the bubble chamber, in
the new context of direct dark matter detection \cite{Bolte:2005}.
We make a case study of it, given that the target liquid employed
(CF$_{3}$I) has an extreme simultaneous sensitivity to
spin-dependent-proton and spin-independent couplings, via the
presence of fluorine \cite{Ellis:1991} and iodine respectively. 
We compare expectations with other
detectors not profiting from this high degree of simultaneous
sensitivity, such as the case of Germanium-based searches. Another
important distinguishing feature of COUPP is the ability to run
modules containing C$_{3}$F$_{8}$ or C$_{4}$F$_{10}$, much more
sensitive to the spin-dependent than -independent contributions to
the signal rate. This multiplicity of targets can be exploited to
identify the nature and properties of a WIMP. The conclusions are,
however, not unique to COUPP and can be extended to argue the
present need for a variety of targets and experiments, if a clear
picture of the characteristics of a dark matter particle is to be
obtained. 

\paragraph{Detection Challenges and COUPP.}
In order to explore extensive regions of the DM parameter space,
direct detection experiments must rise to the challenge of
constructing ton-scale detectors with only a few background events
per year. Even deep underground, where cosmic-ray backgrounds can be
substantially reduced, naturally occurring radioactivity poses quite
a challenge. COUPP (Chicagoland Observatory for Underground Particle
Physics) uses stable room-temperature bubble chambers to search for
DM particles scattering off of nuclei in superheated liquid. The
superheated refrigerant initially used, CF$_3$I, is an inexpensive
fire-extinguishing agent.
CF$_3$I is an excellent WIMP-detector: iodine is an optimal
target for spin-independent (SI) interactions, fluorine is 
the best possible target for spin-dependent-proton (SD$_p$)
interactions, and both iodine and fluorine are good targets for
spin-dependent-neutron (SD$_n$) interactions. Because of COUPP's
simplicity, room temperature operation and the low cost of several
target liquids of interest, COUPP detectors are quite inexpensive as
compared with other approaches to DM detection.

COUPP presently operates a 2 kg chamber at the ~300 meters of water
equivalent depth of the Fermilab neutrino tunnel.
The potential reach of this CF$_3$I-filled chamber at the current
depth is presented in \cite{coupp}.
Which sensitivity is actually achieved will
depend on the level of alpha-emitter contamination in the detection
volume (in contrast to most direct detection experiments, COUPP's
demonstrated minimum ionizing background rejection of $>10^{10}$
makes reduction of alpha-emitters its sole radiopurity concern
\cite{coupp}).

The short-term goals for COUPP are to reduce the alpha-recoil
backgrounds in the 2kg chamber to a level of less than one event per
kg per day, and to apply the upgrades tested on it to larger
chambers currently under construction.  The collaboration is
constructing larger devices, totaling 80 kg of CF$_3$I. Long-term
plans involve the deep-underground installation of a target mass of
order one ton, using a number of different refrigerant targets for
an exhaustive exploration of DM models.  The ability of COUPP to use
the same detector technology to measure interaction rates on a range
of targets is considered to be one of the principal strengths of
this approach.

\paragraph{Theoretical Predictions.}
In SUSY extensions of the SM a discrete
symmetry, known as $R$-parity,
is often imposed in order to forbid lepton and baryon violating
processes which could lead, for instance, to proton decay. A
phenomenological implication of this is that SUSY particles
are only produced or destroyed in pairs, thus rendering the lightest
SUSY particle (LSP) stable. Remarkably, in large areas of
the parameter space of SUSY models, the LSP is an electrically neutral
particle, the lightest neutralino, $\tilde{\chi}^0_1$, which therefore
constitutes a very well motivated DM candidate, within
the class of WIMPs \cite{Jungman:1995df,mireview}.

The neutralino is a linear superposition of the fermionic partners of
the neutral electroweak gauge bosons (bino and wino) and of the
neutral Higgs bosons (Higgsinos),
and the resulting
detection cross section is extremely dependent on its
specific composition.
The scalar
part of the neutralino-proton cross section, $\crosssec$,
receives contributions from Higgs
exchange in a $t$-channel and squark exchange in an $s$-channel. The
latter also contributes to the spin-dependent part of the cross
section, $\crosssecsd$,
together with a $Z$ boson exchange in a $t$-channel.
The expressions for the different amplitudes can be found, e.g., in
\cite{ellis}.
Thus, a large Higgsino component induces an enhancement of
both the Higgs and $Z$ boson exchange diagrams, thereby leading to an
increase in both the spin-dependent and -independent cross sections.
On the other hand, the presence of very light squarks leads to an
enhancement of (mainly) $\crosssecsd$.
\begin{figure*}[t]
  \epsfig{file=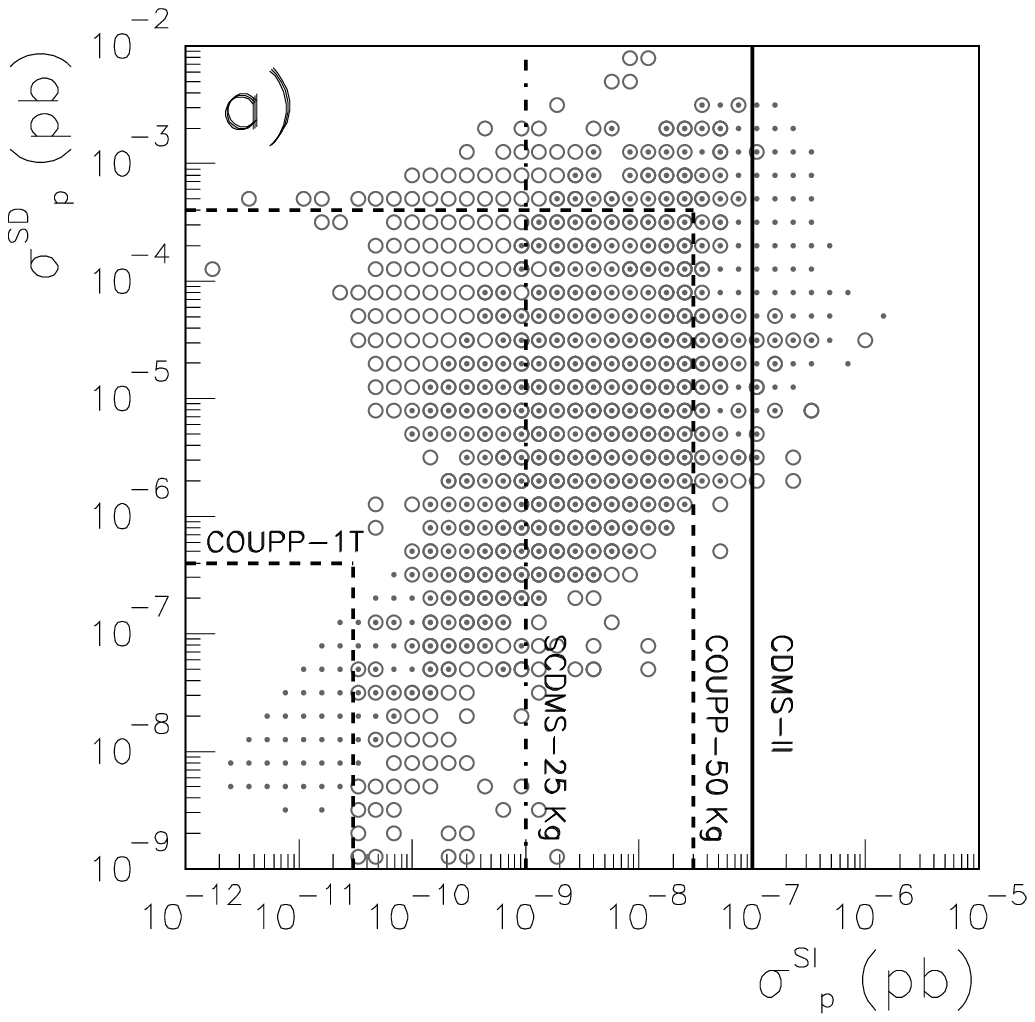, width=0.45\textwidth}
  \epsfig{file=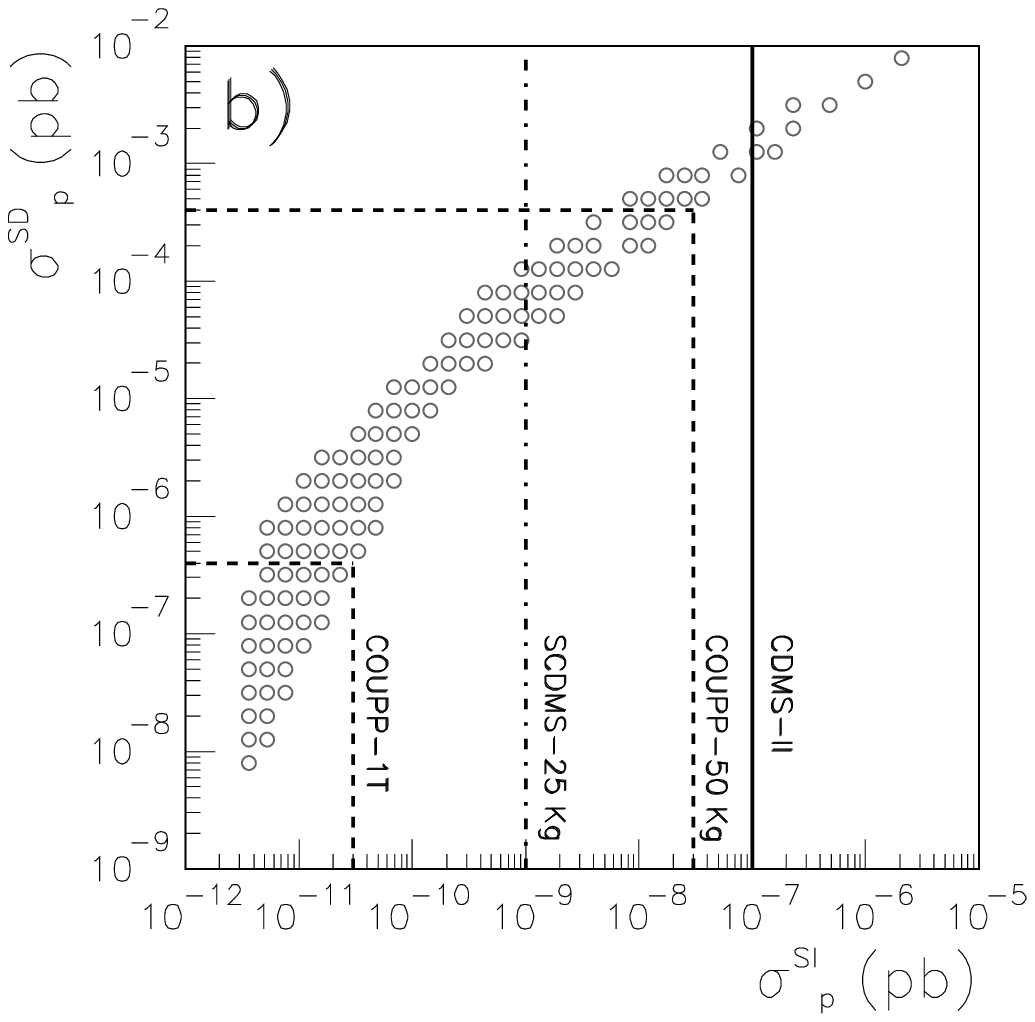,width=0.45\textwidth}
  \vspace*{-0.5cm}
  \caption{ 
    Theoretical predictions for $\ccrosssecsd$ versus
    $\ccrosssec$ obtained from a set of random scans in the various
    supersymmetric (effMSSM and supergravity-inspired)
    scenarios (left) and in the UED scenario (right). All the points
    fulfil existing experimental
    constraints and reproduce the correct dark matter relic density.
    The current and projected sensitivities of the CDMS detector
    (25 kg stage) are also
    represented with solid and dot-dashed lines, respectively,
    together with the potential reach of COUPP (dashed lines). The
    sensitivity of COUPP at 1 ton target mass is based  
    on the goal of matching the lowest alpha-emitter concentrations
    so far achieved in neutrino experiments \protect\cite{coupp} 
    (e.g., KAMLAND \cite{kamland}).}
  \label{effMSSM}
\end{figure*}

In order to determine
the theoretical predictions for the neutralino 
detection cross section we have performed
a random scan in the effective MSSM  (effMSSM)
scenario, where input quantities are defined at the electroweak
scale
\cite{effMSSM}. The mass parameters
have been taken in the range $0\le\mu$, $m_A$, $M_1$,
$m\le 2$~TeV with $3\le\tan\beta\le50$, and $-4\,M_1<A<4\,M_1$ (see
\cite{ellis,bedny} for a similar scan).
A small non-universality in squark soft masses has also been included,
taking $m^2_{Q,u,d}=(1\ldots5)\,m^2$. The results are depicted
in Fig.\,\ref{effMSSM}a) by means of empty circles. Noteworthy,
regions with large $\crosssecsd$ are obtained, some of
which predict a small $\crosssec$.
In a second scan we have studied supergravity-inspired
models in which the soft terms are inputs at the grand
unification scale. We have considered the most general situation,
with non-universal scalar and gaugino masses,
exploring the scenarios presented in
\cite{nunivsugra} for $3\le\tan\beta\le50$
(see for comparison \cite{rosz}, 
where the scenario with universal
parameters is studied).
The results are shown in Fig.\,\ref{effMSSM}a) with
gray dots. 
In this case a simultaneous
increase in both
$\crosssecsd$ and $\crosssec$
is observed.
We leave the details of these scans and the implications on the SUSY
parameter space for a forthcoming work.
These results strongly suggest the need of combining
spin-dependent and -independent techniques in order to effectively
explore the whole SUSY parameter space.

Although theoretically very well motivated, SUSY is not the only
possible extension of the SM leading to a viable DM candidate.
An interesting alternative arises in
theories  with Universal Extra Dimensions (UED), in which all 
fields are allowed to propagate in the bulk \cite{Appelquist:2000nn}.
In this case, the Lightest Kaluza-Klein Particle (LKP) is a
viable DM candidate, likely to be associated with the first KK
excitation  
of the hypercharge gauge boson~\cite{Cheng:2002iz,Servant:2002aq}, 
usually referred to as $\bone$. 
In absence of spectral degeneracies, the $\bone$ would achieve the 
appropriate relic density for masses in the 850--900 GeV
range~\cite{Servant:2002aq}.  
However, due to the quasi-degenerate nature of the KK spectrum, this
range can be significantly
modified, due to coannihilations with
first~\cite{Burnell:2005hm,Kong:2005hn} 
and second~\cite{Kakizaki:2005en,Matsumoto:2005uh,Kakizaki:2006dz}
KK-level modes. The allowed mass range was also found to depend
significantly on  
the mass of the Standard Model Higgs boson~\cite{Kakizaki:2006dz}, and  
in general on the matching contributions to the brane-localized
kinetic terms at the cut-off scale (see the discussion in
Ref.~\cite{Burnell:2005hm}).

Our calculation of the LKP scattering cross section off nucleons
closely follows \cite{Cheng:2002ej} (see also \cite{Servant:2002aq}).
In practice, it is performed in a way very similar to the case of
SUSY,  
evaluating the amplitudes for scattering of the $\bone$ particles off 
nucleons. In 
UED the leading contribution comes from the exchange of the Higgs (for 
scalar 
coupling) and of first level KK quarks $q^{(1)}$ (for both axial and
scalar couplings). We will work under the usual assumption that all
first level KK quarks are degenerate with 
mass $m_{q^{(1)}}$.
The resulting spin-dependent and -independent LKP detection cross
section is represented in Fig.\,\ref{effMSSM}b), where (in view of the  
aforementioned theoretical uncertainties on the $\bone$ parameters)
we took a rather liberal approach,
and let the $\bone$ mass $m_{\bone}$,
and the normalized mass difference between the first level KK quarks
and the $\bone$, $R_{q^{(1)}} \equiv
(m_{\bone}-m_{q^{(1)}})/m_{\bone}$, to vary independently in the 
range $ 300$ GeV $\leq m_{\bone} \leq 2000$ GeV, and $0.01 \leq
R_{q^{(1)}}\leq 0.5$. 
Note that masses $m_{\bone} \lesssim 300$ GeV are excluded by
electroweak 
precision data~\cite{Flacke:2005hb,Gogoladze:2006br}. 
As one can see, LKP models tend to populate a different region of 
the parameter space with respect to SUSY scenarios, due to the
larger spin-dependent cross-section.

\begin{figure*}[t]
  \epsfig{file=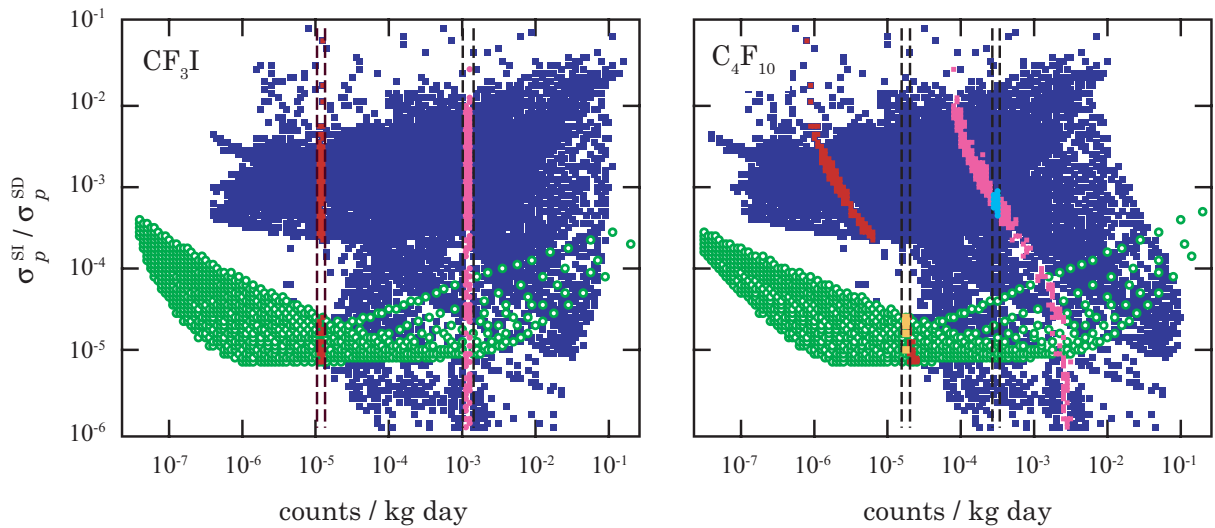,width=0.9\textwidth}
  \vspace*{-0.2cm}
  \caption{
    {\it Left Panel:} The detection of a DM signal with a CF$_3$I
    detector can only loosely constrain DM candidates (blue squares
    for neutralinos, green circles for the LKP) in the
    $\sigma^{\rm SI}_p / \sigma^{\rm SD}_p$ versus count-rate
    plane. Red (magenta) dots show the many models consistent with a
    measurement of  $\sim10^{-5}$ ($10^{-3}$) counts / kg day on
    CF$_3$I. {\it Right Panel:} measurement of the event rate in a
    second detection fluid such as C$_4$F$_{10}$, with lower
    sensitivity to spin-independent couplings, effectively reduces the
    remaining number of allowed models--orange (aqua) dots--and
    generally allows discrimination between the neutralino and the
    LKP (a 10\% uncertainty in the measurements is adopted here for
    illustration).  
  }
  \label{targets}
\end{figure*}

\paragraph{WIMP Discovery and Identification.}
The discovery of neutralino DM might take place 
through either scalar or axial coupling.
In contrast, discovery of
LKP DM is for most, but not all, models expected to occur
through axial coupling. The ability of COUPP to run with a target
such as CF$_3$I, which has optimal SI, SD$_n$, and SD$_p$ couplings,
is an advantage of this experiment in the race for first detection.
Supposing an experiment succeeds in directly detecting DM particles,
it is interesting to consider how the nature of the DM (e.g.
neutralino or LKP) might be determined.  The possibility of running
with a range of detection fluids makes COUPP well-poised
to determine the nature of DM upon successful detection. As shown in
Fig.~\ref{targets}(a), measurement of an event rate in a
single detector does reduce allowed models, but does not generally
place significant constraints on coupling parameters or on the nature
of detected DM (i.e. neutralino or LKP). However, as shown in
Fig.\,\ref{targets}b), subsequent detection of an event rate on a
second target does substantially reduce the allowed range of
coupling parameters, and allows, in most cases, an effective
discrimination 
between
neutralino and LKP DM
(it has recently been pointed out \cite{Hooper:2006xe} that a
combination of direct and indirect detection techniques might also
help distinguishing between these two candidates).
The combination of detector fluids
used in Fig.\,\ref{targets} is effective in reducing the allowed
range of $\sigma^{\rm SI}_p / \sigma^{\rm SD}_p$ because massive
iodine nuclei have a large SI coupling, while fluorine nuclei have a
large SD$_p$ coupling. It must be noted that fluorine and iodine
have very similar neutron cross sections. Monte Carlo simulations
show that CF$_{3}$I and C$_{3}$F$_{8}$ or C$_{4}$F$_{10}$ exhibit
essentially the same response to any residual neutron background,
i.e., neutrons cannot mimic an observed behavior such as that
described in the discussion of Fig.\,\ref{targets}. Other
combinations of targets such as germanium and silicon are
more prone to systematic effects where residual neutron recoils can
mimic the response expected from a WIMP with dominant 
spin-independent couplings.

\paragraph{Conclusions.}
As we have shown  with \frenchspacing{Fig.\,\ref{effMSSM}},
in certain phenomenological scenarios a
detector sensitive exclusively to one mode of interaction may lack
sensitivity to a large fraction of WIMP candidates. The
possibility of operating experiments, such as COUPP, with a range 
of detection fluids, makes them ideally suited to determine the 
nature of dark matter upon
successful detection, i.e., to distinguish between LKP and
neutralino candidates, and in the second case, to pinpoint the
properties of the particle in an otherwise vast supersymmetric
parameter space. The arguments presented here for the case study of
COUPP can be easily generalized to a combination of data from
experiments using targets maximally sensitive to different
couplings, supporting the tenet that a large variety of DM
detection methods is presently desirable.

\paragraph{Acknowledgements.} GB was supported during part of this
project by the Helhmoltz Association of National Research Centres. DGC 
is supported by the program ``Juan de la Cierva'' of the Spanish
Ministry of Science and Education. 
GB and DGC also acknowledge support from the ENTApP Network of the
ILIAS project RII3-CT-2004-506222.
JIC and BO are supported by
the Kavli Institute for Cosmological Physics through grant NSF
PHY-0114422 and by NSF CAREER award PHY-0239812.

\providecommand{\href}[2]{#2}

\end{document}